\documentclass[iop]{emulateapj-rtx4}
\usepackage{apjfonts} 

\usepackage{graphicx}
\usepackage{verbatim}

\newcommand{\Fermi}{{\it{}Fermi}\ }

\makeindex
\citeindextrue

\received{June 9, 2010}
\accepted{August 23, 2010}
\submitted{}

\shorttitle{Radio/gamma-ray time delay in the parsec-scale cores of AGN}
\shortauthors{Pushkarev et al.}

\begin{document}
\title{Radio/gamma-ray time delay in the parsec-scale cores of active 
       galactic nuclei}
\author{
A. B. Pushkarev\altaffilmark{1,2,3},
Y. Y. Kovalev\altaffilmark{4,1},
M. L. Lister\altaffilmark{5}
}
\altaffiltext{1}{
Max-Planck-Institut f\"ur Radioastronomie, Auf dem H\"ugel 69, 
53121 Bonn, Germany;
\email{apushkar@mpifr.de}
}
\altaffiltext{2}{
Pulkovo Observatory, Pulkovskoe Chaussee 65/1, 196140 St. 
Petersburg, Russia;
}
\altaffiltext{3}{
Crimean Astrophysical Observatory, 98409 Nauchny, Crimea, Ukraine;
}
\altaffiltext{4}{
Astro Space Center of Lebedev Physical Institute,
Profsoyuznaya 84/32, 117997 Moscow, Russia;
\email{yyk@asc.rssi.ru}
}
\altaffiltext{5}{
Department of Physics, Purdue University, 525 Northwestern Avenue, 
West Lafayette, IN 47907, USA;
\email{mlister@purdue.edu}
}

\begin{abstract}

We report the detection of a non-zero time delay between radio
emission measured by the VLBA at 15.4~GHz and $\gamma$-ray radiation
($\gamma$-ray leads radio) registered by the Large Area Telescope
(LAT) on board the {\it Fermi Gamma-Ray Space Telescope} for a sample 
of 183 radio and $\gamma$-ray bright active galactic nuclei (AGNs). For 
the correlation analysis we used $0.1-100$~GeV $\gamma$-ray photon
fluxes, taken from monthly binned measurements from the first \Fermi 
LAT catalog, and 15.4 GHz radio flux densities from the MOJAVE VLBA
program. The correlation is most pronounced if the core flux density
is used, strongly indicating that the $\gamma$-ray emission is 
generated within the compact region of the 15~GHz VLBA core. 
Determining the Pearson's $r$ and Kendall's $\tau$ correlation 
coefficients for different time lags, we find that for the majority 
of sources the radio/$\gamma$-ray delay ranges from 1 to 8 months in 
the observer's frame and peaks at approximately 1.2 months in the source's 
frame. We interpret the primary source of the time delay to be  
synchrotron opacity in the nuclear region.
\end{abstract}

\section{INTRODUCTION} 
\label{s:intro}
Long-term systematic observations of active galactic nuclei (AGN) have
produced detailed light curves that are a powerful tool for
investigating the nature of these highly energetic phenomena.  AGNs
show variability across the full electromagnetic spectrum and
multi-frequency monitoring programs serve to establish connections
between flux variations at different energy bands.

The first comparison between long-term records of radio (10.7~GHz) and
optical fluxes for a sample of 24 AGNs \citep{Pomphrey76} showed a
correlation in only 13 sources, with the optical events preceding
radio by intervals of 0 to 14 months. Subsequent analysis of light
curves for 18 AGNs taken in the optical and radio (4.6--14.5~GHz)
domains \citep{Clements95} demonstrated similar statistics: nine
sources exhibited positive radio-optical correlations, with a time lag
ranging from 0 to 14 months. \cite{Tornikoski94} reported on a
correlation in 10 out of 22 sources comparing optical and radio
(22--230~GHz) observations. Remarkably, in 6 sources the variability
was simultaneous. A more recent statistical study of the time delay
between individual millimeter- and centimeter-wave flare peaks made by
\cite{Hovatta_delay} for a sample of 55 sources showed a large scatter of 
time lags that ranged up to several hundreds of days between 4.8 and 
230~GHz, and tended to increase with decreasing radio frequency band. 
\cite{Jorstad10_3C454.3} investigated the flaring behavior of the quasar 
3C~454.3 and showed that optical outbursts led the 230~GHz flares by
15--50 days, confirming an earlier result by \cite{Raiteri08}, who found 
millimeter flux changes lagging behind the optical on a time scale of 
about 60 days. A longer delay of about ten months between optical and 
37~GHz radio flux variations was reported in this source by \cite{Volvach08}. 
\cite{radio-gamma-EGRET} studied the connection between $\gamma$-ray 
emission detected by EGRET and phases of corresponding radio flares at 
22 and 37~GHz. They reported that the highest levels of $\gamma$-ray emission 
were detected 30--70 days after the onset of the high-frequency radio flare.

The {\it{}Fermi Gamma-Ray Space Telescope}, successfully launched in
June 2008, has opened a new era in $\gamma$-ray astronomy. A number of
AGN radio/$\gamma$-ray connections have been established on the basis 
of the first three months of \Fermi science operations and 
quasi-simultaneous VLBI observations, namely that the $\gamma$-ray
photon flux correlates with the parsec scale radio flux density
\citep{MF2}, and that the jets of the LAT-detected blazars have
higher-than-average apparent speeds (median of $15\,c$; 
\citealp{MF1}), larger apparent opening angles (median of
$20^\circ$; \citealp{MF3}), and higher variability Doppler factors
(mean of $20$; \citealp{MF4}). In addition, AGN jets tend to be found
in a more active radio state within several months of the LAT-detection
of their strong $\gamma$-ray emission
\citep{MF2}. A significant correlation was detected even between 
non-simultaneous measurements of $0.1-100$~GeV photon flux and 8~GHz
VLBA flux density \citep{Kovalev_Fermi_assoc}. Nonetheless, there are
several key questions that remain to be addressed, which include: (i)
the dominant production mechanism(s) of $\gamma$-ray photons, (ii) the
exact location of the high-energy emission, and (iii) the nature of
the AGN $\gamma$-ray duty cycle. In this Letter we investigate the
radio/$\gamma$-ray time delays for parsec-scale jets and place
constraints on the localization of the region where most of the
$\gamma$-ray photons are produced.

Throughout this Letter, we use the term ``core'' as the apparent origin
of AGN jets that commonly appears as the brightest feature in VLBI images
of blazars \citep[e.g.,][]{Lobanov98_coreshift,Marscher08}.

\section{THE RADIO DATA AND SOURCE SAMPLE}
\label{s:obs}

The MOJAVE program \citep{MOJAVE} is a long-term VLBA project to study
the total intensity structure changes and polarization evolution of
extragalactic relativistic radio jets in the northern sky. The
observed sources include a statistically complete, flux-density
limited sample of 135 AGNs (MOJAVE-1). All the MOJAVE-1 sources have
J2000 declination $\delta>-20\degr$ and a 15~GHz VLBA correlated flux
density $S_\mathrm{corr}>1.5$~Jy (2~Jy for $\delta<0\degr$) at any
epoch between 1994.0 and 2004.0. The weaker radio blazars
($S_\mathrm{corr}>0.2$~Jy) detected by \Fermi extend the complete
MOJAVE-1 sample to MOJAVE-2. The monitoring list currently consists 
of 293 sources, 186 of which are members of the First \Fermi LAT catalog
\citep[1FGL,][]{1FGL} that are positionally associated with AGNs.
We note that the 186 sources do not represent a complete sample selected on 
either parsec-scale radio flux density or $\gamma$-ray photon flux.

\section{RESULTS}
\label{s:results}
\subsection{Radio/$\gamma$-ray delay in VLBI cores}
\label{s:rg_core}

Apart from the median $\gamma$-ray photon and energy fluxes, the 1FGL
catalog provides flux history data, in the form of monthly binned
$0.1-100$~GeV photon flux measurements during the first 11 months
of the \Fermi scientific operations, which started on 2008 August
4. The time sampling of our VLBA radio observations is
source-dependent: objects with more rapid structural changes (i.e., faster
apparent speeds) are observed more frequently. There are only five
sources in our sample that are monitored with a cadence more frequent 
than once every two months. Starting in early 2009, fifty five bright
$\gamma$-ray detections ($>10\sigma$) from the \Fermi LAT 3-month list
positionally associated with bright radio-loud blazars
\citep{Fermi3ml,LBAS,Kovalev_Fermi_assoc} have been incrementally added
to the MOJAVE program. More than half of these new LAT-detected sources have
fewer than three epochs of radio observations during the \Fermi era,
which precludes the correlation analysis of individual light
curves. Therefore, our study is based upon a statistical approach.

Overall, we obtained 564 VLBA images and corresponding model fits for 183 
bright $\gamma$-ray sources (Table~\ref{t:sample}) within a period from June
2008 through March 2010. The parsec-scale structure, typically represented 
by a one-sided core-jet morphology, was fitted with the procedure {\it modelfit} 
in the Difmap package \citep{difmap} using a limited number of primarily 
circular Gaussian components, as described by \cite{MOJAVE}.

\begin{figure*}[t]
\begin{center}
\includegraphics[width=9.4cm,angle=-90]{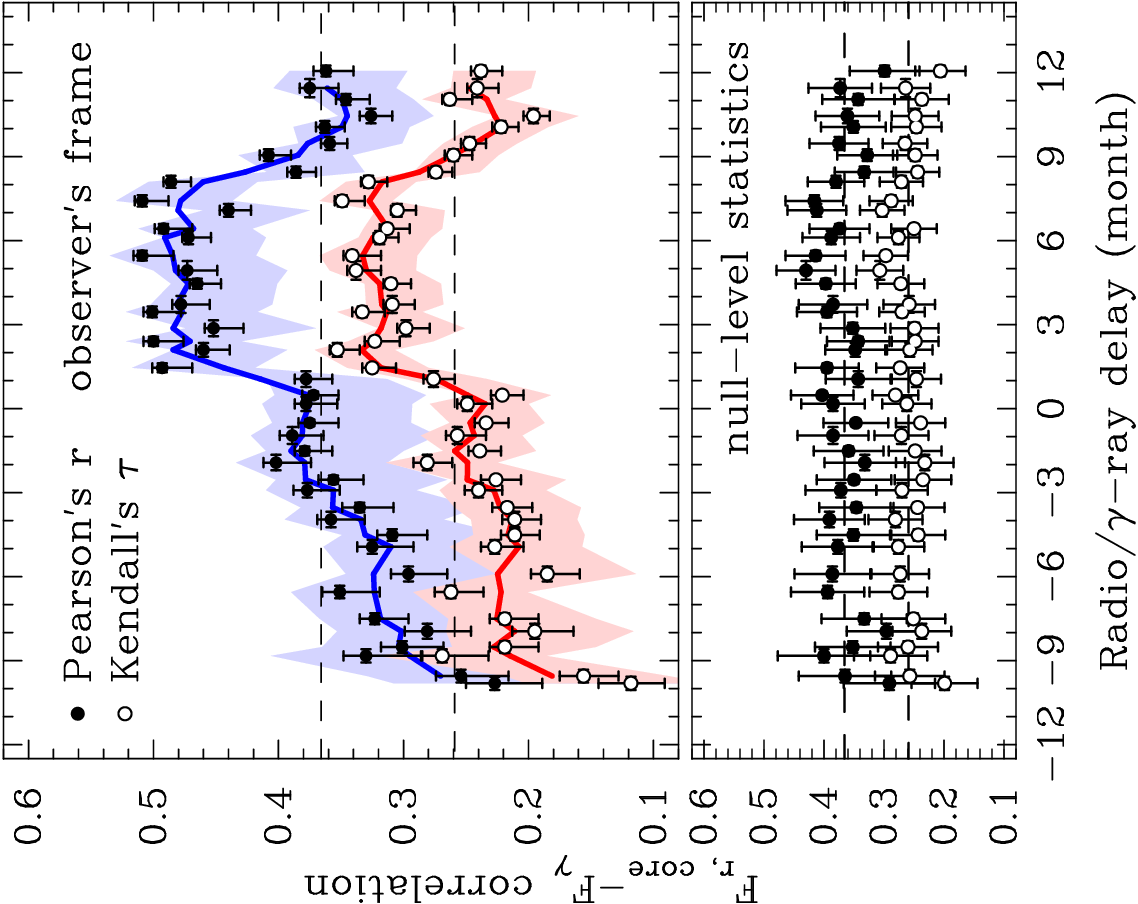}\hspace{1cm}
\includegraphics[width=9.4cm,angle=-90]{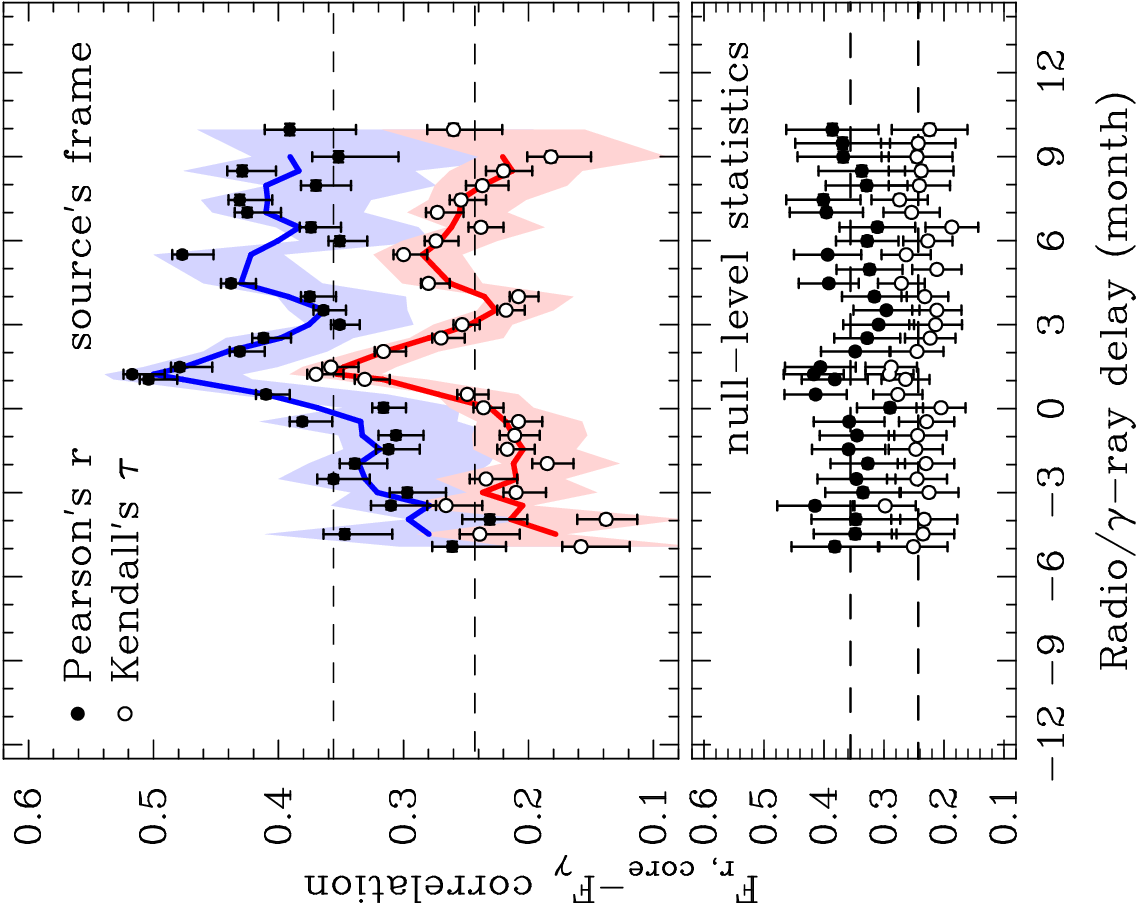}
\end{center}
\caption{\label{f:gr_delay}
{\it Left:} Upper panel: flux-flux correlation level as a function of
the time interval between integrated $0.1-100$~GeV photon flux and
radio core flux density measurements in the observer's frame. Filled
circles represent the Pearson's $r$ statistic, and open circles denote
the Kendall's $\tau$ statistic. The error bars are shown at the 68\%
level, while shaded areas represent 95\% confidence regions. The thick 
curves are constructed by applying a three point moving average. 
Dashed lines show the null-basis level of the correlation due to the 
overall $\gamma$-ray photon flux versus radio flux density correlation. 
Lower panel: estimation of the null level correlation obtained by 
shuffling $\gamma$-ray photon fluxes for every source and keeping 
radio flux densities the same. {\it Right:} same curves, corrected 
to the source rest frame. A sharp peak at a delay of $\sim1.2$~months 
is detected.}
\end{figure*}

We tested for possible correlations between the $\gamma$-ray photon
fluxes and 15~GHz VLBA core flux densities using the following
procedure: (i) we selected all pairs of measurements where the
difference in the radio and $\gamma$-ray epochs lay within a
restricted time interval, for instance, $[-0.5,+0.5]$ month, 
where the negative sign indicates that the radio measurement precedes 
the $\gamma$-ray one, (ii) if  
more than one pair of fluxes was available for a source, we selected
the one with the epoch difference closest to the mean of the time
interval. The procedure was then repeated iteratively by shifting the
time interval by 0.5~month each time. We performed a quantative analysis 
that confirmed that our data do not provide any bias towards positive pairs 
of radio/$\gamma$-ray epoch difference. We used a cutoff of
$\mathrm{SNR}>3$ for the $\gamma$-ray photon flux measurements
to avoid a bias due to the lack of sources that are both
weak in radio and $\gamma$-rays, and to exclude the influence of
low-quality data points. For each data set we calculated the Pearson's
$r$ and non-parametric Kendall's $\tau$ correlation coefficients,
together with a corresponding probability of a chance correlation
(Table~\ref{t:corr_stat}). A non-zero radio/$\gamma$-ray time lag,
clearly seen as a bump in the correlation versus delay curves
(Fig.~\ref{f:gr_delay}, left top panel), ranges from 1 to 8
months. The smooth fitted curves were obtained by applying a three
point moving average.

\begin{deluxetable}{ccccc}
\tablecolumns{5}
\tablewidth{0pc}
\tabletypesize{\scriptsize}
\tablewidth{0pt}
\tablecaption{\label{t:sample}Sample of LAT-detected MOJAVE AGNs}
\tablehead{
\colhead{Source}     & \colhead{Alias} &
\colhead{\Fermi name} & \colhead{$z$}  &
\colhead{Redshift reference}\\
(1) & (2) & (3) & (4) & (5)}
\startdata
$0003+380$  &            \nodata  & 1FGL J$0005.7+3815$  &  0.229 & \cite{Fermi_AGN_1yr}\\
$0015-054$  &            \nodata  & 1FGL J$0017.4-0510$  &  0.227 & \cite{Fermi_AGN_1yr}\\
$0048-071$  &          OB $-$082  & 1FGL J$0051.1-0649$  &  1.975 & \cite{Fermi_AGN_1yr}\\
$0048-097$  &            \nodata  & 1FGL J$0050.6-0928$  &\nodata & \nodata\\
$0059+581$  &            \nodata  & 1FGL J$0102.8+5827$  &  0.644 & \cite{Fermi_AGN_1yr}\\
$0106+013$  &          4C +01.02  & 1FGL J$0108.6+0135$  &  2.099 & \cite{Fermi_AGN_1yr}\\
$0106+678$  &          4C +67.04  & 1FGL J$0110.0+6806$  &  0.290 & \cite{Fermi_AGN_1yr}\\
$0109+224$  &            \nodata  & 1FGL J$0112.0+2247$  &  0.265 & \cite{Fermi_AGN_1yr}\\
$0110+318$  &          4C +31.03  & 1FGL J$0112.9+3207$  &  0.603 & \cite{Fermi_AGN_1yr}\\
$0113-118$  &            \nodata  & 1FGL J$0115.5-1132$  &  0.672 & \cite{Fermi_AGN_1yr}
\enddata
\tablecomments{Columns are as follows:
(1) IAU name (1950);
(2) alternate name;
(3) 1FGL name;
(4) redshift;
(5) literature reference for redshift.
Table~\ref{t:sample} is published in its entirety in the electronic version of the 
{\it Astrophysical Journal}. A portion is shown here for guidance.}
\end{deluxetable}

To test the robustness of this result, we estimated the uncertainty
value of the correlation coefficients. Since both the radio flux
densities and the $\gamma$-ray photon fluxes are far from being
normally distributed, direct methods like the Fisher transformation or
Student's t-distribution could not be used. Thus, we applied
randomization techniques based on permutation tests to construct
confidence intervals on the correlation coefficients. For each data
set the randomization was done in the following manner: (i) we
randomly swapped the radio measurements for one source with another
source, keeping the $\gamma$-ray fluxes the same; (ii) we calculated
correlation coefficients $r$ and $\tau$ from the randomized data. We
then repeated these steps 2000 times. A 95\% confidence interval
(given in Table~\ref{t:corr_stat}) for the correlation coefficients
was defined as the interval spanning from the 2.5-th to the 97.5-th
percentile of the re-sampled $r$ and $\tau$ values. To estimate the
null-basis level of the flux-flux correlation (which is present due to
an overall radio/$\gamma$-ray correlation; see \S~\ref{localization})
, we shuffled the $\gamma$-ray photon fluxes among the 11 measurements
available for every source, keeping the epoch dates and radio flux
densities the same. The resulting values were $r_0=0.37$ and
$\tau_0=0.26$ (Fig.~\ref{f:gr_delay}, left bottom panel). When we
randomly selected 90\% (and 80\%) of the sample the correlations
remained significant, indicating that they are not driven by
outliers. Additionally, we found no significant correlation between
redshift and the VLBA core flux density averaged for the sample over
the \Fermi era.

\begin{deluxetable}{rrccc}
\tablecolumns{5}
\tabletypesize{\scriptsize}
\tablewidth{0pt}
\tablecaption{\label{t:corr_stat}Radio/$\gamma$-ray flux-flux correlation statistics for the VLBA cores in the source frame}
\tablehead{
\colhead{$\Delta t_{\mathrm{R}-\gamma}$} & \colhead{$N$} &
\colhead{$r$} & \colhead{$\tau$} & \colhead{$p$} \\
(1)~~~~~ & (2) & (3) & (4) & (5)}
\startdata
$-4.9 \pm 0.2$ &  38  & $0.26^{+0.04}_{-0.20}$ & $0.16^{+0.06}_{-0.09}$ & $1.6\times 10^{-1}$ \\
$-4.5 \pm 0.2$ &  50  & $0.35^{+0.06}_{-0.11}$ & $0.24^{+0.05}_{-0.07}$ & $1.4\times 10^{-2}$ \\
$-4.0 \pm 0.2$ &  42  & $0.23^{+0.07}_{-0.11}$ & $0.14^{+0.06}_{-0.06}$ & $2.0\times 10^{-1}$ \\
$-3.5 \pm 0.1$ &  54  & $0.31^{+0.05}_{-0.11}$ & $0.27^{+0.04}_{-0.07}$ & $4.5\times 10^{-3}$ \\
$-3.0 \pm 0.2$ &  59  & $0.30^{+0.04}_{-0.10}$ & $0.21^{+0.04}_{-0.07}$ & $1.9\times 10^{-2}$ \\
$-2.5 \pm 0.1$ &  58  & $0.36^{+0.04}_{-0.11}$ & $0.23^{+0.04}_{-0.06}$ & $9.4\times 10^{-3}$ \\
$-2.0 \pm 0.2$ &  60  & $0.34^{+0.04}_{-0.11}$ & $0.18^{+0.04}_{-0.06}$ & $3.6\times 10^{-2}$ \\
$-1.5 \pm 0.1$ &  70  & $0.31^{+0.04}_{-0.09}$ & $0.22^{+0.03}_{-0.06}$ & $8.0\times 10^{-3}$ \\
$-1.0 \pm 0.2$ &  68  & $0.31^{+0.05}_{-0.08}$ & $0.21^{+0.03}_{-0.06}$ & $1.1\times 10^{-2}$ \\
$-0.5 \pm 0.1$ &  76  & $0.38^{+0.03}_{-0.12}$ & $0.21^{+0.03}_{-0.05}$ & $7.9\times 10^{-3}$ \\
$ 0.0 \pm 0.2$ &  90  & $0.32^{+0.03}_{-0.06}$ & $0.24^{+0.03}_{-0.04}$ & $9.6\times 10^{-4}$ \\
$ 0.5 \pm 0.1$ &  95  & $0.41^{+0.03}_{-0.10}$ & $0.25^{+0.03}_{-0.04}$ & $3.5\times 10^{-4}$ \\
$ 1.0 \pm 0.1$ &  85  & $0.50^{+0.03}_{-0.08}$ & $0.33^{+0.02}_{-0.05}$ & $7.3\times 10^{-6}$ \\
$ 1.2 \pm 0.1$ &  87  & $0.52^{+0.02}_{-0.09}$ & $0.37^{+0.02}_{-0.05}$ & $4.0\times 10^{-6}$ \\
$ 1.5 \pm 0.1$ &  83  & $0.48^{+0.03}_{-0.08}$ & $0.36^{+0.02}_{-0.05}$ & $1.6\times 10^{-6}$ \\
$ 2.0 \pm 0.1$ &  91  & $0.43^{+0.03}_{-0.07}$ & $0.32^{+0.02}_{-0.04}$ & $9.0\times 10^{-6}$ \\
$ 2.5 \pm 0.1$ &  89  & $0.41^{+0.03}_{-0.07}$ & $0.27^{+0.03}_{-0.05}$ & $1.8\times 10^{-4}$ \\
$ 3.0 \pm 0.1$ &  96  & $0.35^{+0.03}_{-0.06}$ & $0.25^{+0.02}_{-0.04}$ & $2.6\times 10^{-4}$ \\
$ 3.5 \pm 0.1$ & 100  & $0.36^{+0.03}_{-0.07}$ & $0.22^{+0.02}_{-0.04}$ & $1.3\times 10^{-3}$ \\
$ 4.0 \pm 0.2$ &  98  & $0.38^{+0.03}_{-0.08}$ & $0.21^{+0.02}_{-0.04}$ & $2.4\times 10^{-3}$ \\
$ 4.5 \pm 0.1$ &  99  & $0.44^{+0.02}_{-0.07}$ & $0.28^{+0.02}_{-0.04}$ & $4.1\times 10^{-5}$ \\
$ 5.5 \pm 0.1$ &  86  & $0.48^{+0.02}_{-0.08}$ & $0.30^{+0.02}_{-0.05}$ & $4.4\times 10^{-5}$ \\
$ 6.0 \pm 0.2$ &  82  & $0.35^{+0.03}_{-0.07}$ & $0.27^{+0.03}_{-0.05}$ & $2.6\times 10^{-4}$ \\
$ 6.5 \pm 0.2$ &  77  & $0.37^{+0.03}_{-0.09}$ & $0.24^{+0.03}_{-0.05}$ & $2.2\times 10^{-3}$ \\
$ 7.0 \pm 0.2$ &  74  & $0.42^{+0.03}_{-0.09}$ & $0.27^{+0.02}_{-0.06}$ & $5.8\times 10^{-4}$ \\
$ 7.5 \pm 0.2$ &  72  & $0.43^{+0.03}_{-0.10}$ & $0.25^{+0.03}_{-0.05}$ & $1.6\times 10^{-3}$ \\
$ 8.0 \pm 0.2$ &  63  & $0.37^{+0.04}_{-0.10}$ & $0.24^{+0.04}_{-0.07}$ & $6.0\times 10^{-3}$ \\
$ 8.5 \pm 0.2$ &  58  & $0.43^{+0.05}_{-0.14}$ & $0.22^{+0.04}_{-0.06}$ & $1.5\times 10^{-2}$ \\
$ 9.0 \pm 0.2$ &  42  & $0.35^{+0.07}_{-0.16}$ & $0.18^{+0.06}_{-0.09}$ & $8.9\times 10^{-2}$ \\
$10.0 \pm 0.2$ &  38  & $0.39^{+0.07}_{-0.20}$ & $0.26^{+0.06}_{-0.10}$ & $2.1\times 10^{-2}$ \\
\enddata
\tablecomments{Columns are as follows:
(1) radio-gamma time delay bin in months;
(2) number of AGNs included in the bin;
(3) Pearson's $r$ with uncertainties at a 95\% confidence level;
(4) Kendall's $\tau$ with uncertainties at a 95\% confidence level;
(5) probability of a chance correlation.}
\end{deluxetable}

The wide range of delays in which the flux-flux correlations are
significant (Fig \ref{f:gr_delay}, left) is presumably a result of the
multiple parameters that determine the conditions in the nucleus
(black hole mass, its spin and accretion rate), and in the nearby
interstellar medium. The delay is also affected by geometry, including
the angle of the jets to our line of sight and the wide range of
redshifts in our sample. The redshifts are known for more than 90\% of
the sources (166 out of 183). We re-did the analysis in the source's
frame by dividing the radio/$\gamma$-ray epoch time difference for
each source by a factor of $(1+z)$. This gave a typical time delay of
$\sim1.2$ months in the source frame (Fig. \ref{f:gr_delay}, right),
which corresponds to $\sim2.5$ months (for $z\sim1$) in the observer's
frame. The other sub-peaks are not significantly different from the
null level of correlation, though they may indicate longer delays in a
smaller number of sources. Note that the points on the
radio/$\gamma$-ray correlation curve are dependent. We expect the
smearing even in the source frame delay because the core size is
redshift dependent: more distant sources have higher rest-frame
frequencies and, therefore should have a smaller core radius and
shorter rest-frame delay. This effect can potentially be studied in
more detail using a larger sample subdivided into high- and
low-redshift bins. Since a distribution of different delay values
is expected, we do not estimate an error range on the detected
peak. The peak value should only be taken as a typical one for the
AGNs in our sample.

\begin{figure}[t]
\begin{center}
\resizebox{0.9\hsize}{!}{\includegraphics[angle=-90]{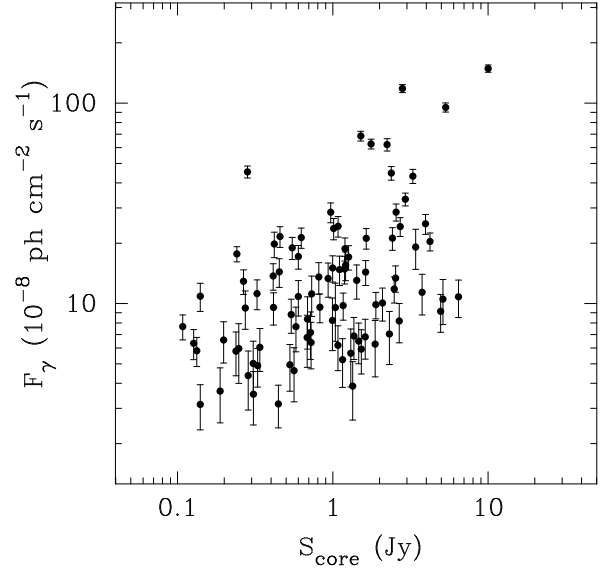}}
\end{center}
\caption{\label{f:flux-flux}
Integrated $0.1-100$~GeV \Fermi LAT photon flux versus
15~GHz VLBA core flux density for data pairs in which the VLBA flux
density measurement was taken $2.5 \pm 0.2$ months after the LAT flux
measurements. One point is plotted per AGN.}
\end{figure}

In Figure~\ref{f:flux-flux} we plot the integrated $0.1-100$~GeV
$\gamma$-ray photon flux against the 15~GHz VLBA core flux density for
data pairs in which the $\gamma$-ray measurement leads the radio
measurement by $2.5 \pm 0.2$ months in the observer's frame. The
formal probability of a chance correlation is $5\times10^{-6}$.  If we
drop the points with photon flux greater than
$4\times10^{-7}$~ph~cm$^{-2}$~s$^{-1}$, the correlation is still
present at a very high level of significance. We also note the span
over one order of magnitude in the $\gamma$-ray fluxes for a given
radio flux density. Several factors can contribute to this scatter:
the shape of the $\gamma$-ray spectrum, K-correction effects,
different seed photons, and different Doppler boosting levels in the
$\gamma$-ray and radio domains \citep{Lister_MC}.

\subsection{\label{localization}Localization of the $\gamma$-ray emission region}

Although radio loud AGN are generally known to have twin jet
structures, the sources in our sample typically have a one-sided
parsec-scale morphology \citep{MOJAVE}, implying strong selection effects
and Doppler boosting of the jet emission.  Localization of the
$\gamma$-ray emission region within the AGN radio structure and its physical production
mechanism(s) remain topics of active debate.  There are three
possibilities for the site of the high-energy emission: (a) only
within the unresolved radio core, (b) only in the resolved
(downstream) jet region, (c) both in the core and jet.

Repeating the same analysis as in \S~\ref{s:rg_core} for the total
VLBA flux density, we found that the corresponding correlation
coefficients agree within the 95\% errors with those found for the
core flux densities. Thus, no firm conclusions can be drawn out of this
comparison. Indeed, the sources in our sample are highly core
dominated, with a median value of
$S_\mathrm{core}/S_\mathrm{VLBA}=0.71$, where $S_\mathrm{core}$ and
$S_\mathrm{VLBA}$ are the core and total VLBA flux density,
respectively. By contrast, when the jet flux densities
($S_\mathrm{VLBA}-S_\mathrm{core}$) are used, the correlation
coefficients are significantly lower, making scenario (b) less
probable. To reduce the uncertainty in the jet flux density estimations
we excluded sources with a high ($>0.9$) core dominance. Nevertheless, 
the correlation between the jet flux densities
measured quasi-simultaneously with the $\gamma$-ray photon flux is
still significant. This could be the result of both radio and $\gamma$-ray 
regimes being boosted by similar beaming factors \citep{MF1,MF2,MF4}. 
Assuming the same Doppler factor for the jet and the core (opaque jet base), 
we would expect a weak but significant correlation between the jet flux
density and integral $\gamma$-ray emission. Alternatively, it is
possible that for some low-redshift sources, $\gamma$-ray emission may 
also be produced outside the radio core, as could be the case in
3C\,84 \citep{3C84_Fermi}, where radio flare accompanying the
$\gamma$-ray activity was detected in the innermost jet region.

We note that \cite{Kovalev_Fermi_assoc} found a correlation between
the \Fermi LAT $0.1-100$~GeV photon flux from the three-month
integration and 8~GHz radio flux density {\it non-simultaneously}
measured by the VLBA. The presence of this correlation in this highly
variable population further suggests that Doppler beaming is the
likely cause, and that the Doppler factors of the individual jets are
not changing substantially over time.

\section{DISCUSSION}
What is the main source of the detected time lag? Dispersion in the
intervening medium cannot be responsible for a delay of several months. 
Our calculations following the Crab pulsar measurement by \cite{Crab_Fermi} 
show that for an object at a redshift $z=1$ the delay should not exceed a few
seconds at 15~GHz.

The most likely source of the observed time lag is synchrotron opacity in 
the nuclear region. The radio core is optically thick to synchrotron emission 
up to the frequency-dependent radius $r_\mathrm{c}\propto\nu^{-1}$ 
\citep{BlandfordKonigl79}. This means that the $\gamma$-ray production 
region/zone is located upstream with respect to the 15~GHz apparent core
 position. Being induced by the same disturbance at a distance $r_\gamma$ 
from the black hole, the radio and $\gamma$-ray emission at their peaks are 
not observed simultaneously due to the opacity effect. Although the 
$\gamma$-ray photons escape immediately, it takes several more months for 
the perturbation to propagate farther along the jet until it reaches the 
$\tau\simeq1$ surface at 15~GHz radio emission (the radio core), and
becomes detectable at radio frequencies. The corresponding distance travelled
along the jet between the place where a $\gamma$-ray photon was emitted 
($r_\gamma$) and the radius of the radio core ($r_\mathrm{c}$) is
\begin{equation}
\Delta r = r_\mathrm{c}-r_\gamma=\frac{\delta\Gamma\beta c\Delta t_{\mathrm{R}-\gamma}^{\mathrm{obs}}}{(1+z)}\,,
\end{equation}
where $\delta$ is the Doppler factor, $\Gamma$ is the Lorentz factor, $\beta c$ 
is the intrinsic jet speed, and $\Delta t_{\mathrm{R}-\gamma}^\mathrm{obs}$ is 
the observed time delay. Using the relations for Doppler factor $\delta$ and 
apparent angular speed $\beta_\mathrm{app}$ \citep{Cohen07}, the expression (1) 
can also be written in the form
\begin{equation}
\Delta r = \frac{\beta_\mathrm{app}c\Delta t_{\mathrm{R}-\gamma}^\mathrm{sour}}{\sin\theta}\,,
\end{equation}
where $\beta_\mathrm{app}$ is the apparent jet speed, 
$\Delta t_{\mathrm{R}-\gamma}^\mathrm{sour}$ is the radio to $\gamma$-ray 
time delay in the source's frame, and $\theta$ is the viewing angle. Since 
we cannot measure radio/$\gamma$-ray time delays for the sources individually
from our data, let us consider a source with a set of parameters typical for 
a LAT-detected radio-loud blazar: apparent jet speed 
$\beta_\mathrm{app}\sim15$ \citep{MF1}, viewing angle $\theta\sim3.6^\circ$ 
\citep{MF3,Hovatta09}, and a typical time lag of 1.2 months in the source's 
frame. Under these assumptions we obtain the distance between the $\gamma$-ray 
production zone and the $\tau\approx1$ surface at 15~GHz, the radio core, 
$\Delta r\sim7$~pc, which corresponds to a projected distance of $\sim0.9$~pc, 
or $\sim0.1$~mas  for a source at $z\sim1$. These estimates are consistent with 
the core radius obtained from the frequency dependent core shift measurements 
\citep{Lobanov98_coreshift,Kovalev08_coreshift_EVN,Kovalev08_coreshift_Crete,Sullivan09_coreshift}.
This indicates that VLBI observations at higher frequencies, for instance at 
43~GHz and 86~GHz, should register shorter delay or even quasi-simultaneous 
flux variations with $\gamma$-ray flux at least for some sources, and might 
even resolve the region of the jet where $\gamma$-ray emission is generated.

The fact that the radio emitting region may be more physically
extended than the $\gamma$-ray emission zone may also contribute to
the detected delay. Finally, we note that our analysis is sensitive to
the delay between the peaks at radio and $\gamma$-ray frequencies, but
cannot address which flare, low- or high-energy, originates first. The
latter analysis requires very well-sampled light curves for a
large set of $\gamma$-ray-detected AGN.

\section{SUMMARY}
\label{s:sum}

We have investigated the dependence between the integrated $0.1-100$~GeV
$\gamma$-ray photon flux and 15~GHz radio flux densities for a large sample 
of 183 LAT-detected AGNs observed by the VLBA within the MOJAVE program and 
conclude the following. 

1. The correlation between $\gamma$-ray photon flux and radio flux
   density is found to be highly significant. The correlation analysis
   results for the core and total VLBA flux density are indistinguishable. 
   The correlation is systematically weaker if the jet flux density is used, 
   providing further support for the localization of the $\gamma$-ray emission 
   to the core region as well as a connection between the inverse-Compton 
   $\gamma$-rays and the synchrotron radio emission from the jet.

2. We found a non-zero radio/$\gamma$-ray delay ($\gamma$-ray leads the radio 
   emission) that ranges from 1 to 8 months in the observer's frame and 
   peaks at $\sim1.2$ months in the source frame. The delay is most likely
   connected with synchrotron opacity in the core region, although other 
   mechanisms may play a role.

3. The region where most of $\gamma$-ray photons are produced is found to be 
   located within the compact opaque parsec-scale core.
   The typical distance between the $\gamma$-ray production region and the 
   15~GHz radio core is estimated to be $\sim7$~pc, which is consistent with 
   the typical core radius derived from frequency dependent core shift 
   measurements.

These results are consistent with earlier findings reported by \cite{MF2} 
and \cite{Tavecchio10} which place further constraints on the localization 
of the $\gamma$-ray production zone in parsec-scale jets. The MOJAVE program is 
continuing to monitor a large sample of LAT-detected AGNs for use in more 
comprehensive studies associated with the next \Fermi data release.

\acknowledgments
We thank M.~H.~Cohen, K.~I.~Kellermann, E.~Ros, T.~Savolainen, and the rest of
the MOJAVE team for useful discussions.
We thank the anonymous referee as well as
A. Wehrle, T. Readhead, F. Tavecchio, and E. Valtaoja
for useful comments.
This research has made use of data from the MOJAVE database that is maintained 
by the MOJAVE team \citep{MOJAVE}. The MOJAVE project is supported under NSF 
grant AST-0807860 and NASA \Fermi grant NNX08AV67G.
Y.~Y.~Kovalev was supported in part by the return fellowship of Alexander 
von Humboldt foundation and the Russian Foundation for Basic Research grant 
08-02-00545.
\facility[NRAO(VLBA)]{The VLBA is a facility of the National Science
Foundation operated by the National Radio Astronomy Observatory under
cooperative agreement with Associated Universities, Inc.}

{\it Facilities:} \facility{VLBA, \Fermi(LAT)}.

\end{document}